\documentclass[aps,notitlepage,oneside,superscriptaddress,11pt]{revtex4-1}
\usepackage{graphicx}
\usepackage{latexsym}
\usepackage{amssymb}
\usepackage{amsmath}
\usepackage{float}
\usepackage{pxfonts}
\usepackage{xcolor}
\definecolor{mygreen}{cmyk}{0.82,0.11,1,0.25}
\definecolor{myyellow}{RGB}{0.4,0.8,0.1}

\begin{document}
\title{Spins of supermassive black holes M87* and SgrA* \\ revealed from the size of dark spots in Event Horizon Telescope Images} 
\author{Vyacheslav Ivanovich Dokuchaev}\thanks{dokuchaev@inr.ac.ru}
\affiliation{Institute for Nuclear Research of the Russian Academy of Sciences, prospekt 60-letiya Oktyabrya 7a, Moscow 117312, Russia}


\begin{abstract}
We reconstructed dark spots in the images of supermassive black holes SgrA* and M87* provided by the Event Horizon Telescope (EHT) collaboration by using the geometrically thin accretion disk model. In this model, the black hole is highlighted by the hot accretion	matter up to the very vicinity of the black hole event horizon. The existence of hot accretion matter in the vicinity of black hole event horizons is predicted by the Blandford--Znajek mechanism, which is confirmed by recent general relativistic MHD simulations in supercomputers. A dark spot in the black hole image in the described model is a gravitationally lensed image of an event horizon globe. The lensed images of event horizons are always projected at the celestial sphere inside the awaited positions of the classical black hole shadows, which are invisible in both cases of M87* and SgrA*. We used the sizes of dark spots in the images of SgrA* and M87* for inferring their spins, $0.65<a<0.9$ and $a>0.75$, accordingly.
\end{abstract}
\keywords{General relativity \and Black holes \and Event horizon \and Gravitational lensing}
\maketitle 

\section{Introduction}

The first images of Supermassive Black Holes (SMBHs) M87* \cite{EHT1,EHT2,EHT3,EHT4,EHT5,EHT6} and SgrA* \cite{EHT1S,EHT2S,EHT3S,EHT4S,EHT5S,EHT6S}, obtained by the Event Horizon Telescope (EHT) collaboration, opened the road for the verification of modified gravity. It must be stressed that this is a unique possibility for an experimental verification of modified gravity  theories. The strong field limit means the applicability of gravity theory at the limiting values of the Newtonian gravitational potential $GM/r\simeq c^2$, where $G$ is the Newtonian gravitational constant, $M$ is the mass of the gravitating object, $r$ is the characteristic radius of the gravitating object, and $c$ is the velocity of light. The most intriguing features of EHT images are the sizes and forms of the dark spots and the surrounding luminous rings with bright spots. 

It is natural to suppose that luminous rings are produced by the emission of hot accretion plasma. The radii of luminous rings in both EHT images very nearly coincide with the corresponding sizes of classical black hole shadows \cite{Bardeen73,Falcke00,ZakhPaoIngrNuc05,JohPsaltis10,Grenzebach14,Grenzebach15,Gralla15,Strom16,Gralla16,Strom17,Strom18,Cunha18a,Cunha18b,Huang18,BisnovatyiTsupko18}. Meanwhile, the sizes of dark spots in both EHT images are suitably smaller than the corresponding black hole shadows; see Equation (\ref{A10}) in Appendix~\ref{AppendixA} for a useful formal expression of the classical black hole shadow. Some examples of the classical black hole shadows are shown in Figure~\ref{fig9} for different spin parameters $a$ in the Kerr metric and possible inclinations of SMBHs M87* and SgrA* with respect to a distant observer line of sight.

There are at least five different methods known to contain spins of SMBHs: thermal continuum fitting, disk reflection modeling, high-frequency quasi-periodic oscillations, X-ray polarimetry, and using the EHT images (see, e.g., the review by {Laura Brenneman 2013}  
\cite{Brenneman} or Christopher S. Reynolds  \cite{Reynolds}). Measuring spins with the EHT images is only suitable for SMBHs Sgr A* and M87*.


Recent magnetohydrodynamic numerical simulations in the framework of General Relativity (GRMHD) (for representative examples, see  \cite{Tchekhovskoy11,Tchekhovskoy12,Tchekhovskoy12b,Tchekhovskoy15,Tchekhovskoy17,Tchekhovskoy17b,Ryan18}) confirm the Blandford--Znajek mechanism \cite{BlandfordZ} of extracting energy from rotating black holes and generating the relativistic jets. In this mechanism, the regular quasi-stationary toroidal and poloidal magnetic fields are generated in the accretion plasma with electric currents penetrating inside the rotating black hole. In turn, the accretion plasma, heated by these electric currents, is very hot and luminous very near to the black hole  event horizon. As a result, the dark spots in the EHT images may be the lensed dark silhouettes of the corresponding event horizon globes, illuminated (highlighted) by the hot accretion plasma \cite{Luminet79,Dexter09,Bromley97,Fanton97,Armitage03,Fukue03,Fukue03b,Tamburini11,Ru-SenLu16,Luminet19}. In the case of a thin accretion disk, the corresponding outline of the dark spot is the gravitationally lensed image of the event black hole horizon equator (see details in  \cite{Dokuch22,doknaz21,doknaz20,Dokuch19,doknaz19,doknazsm19}).

Here, we reconstructed the  form of a dark spot in the black hole image by using numerical calculations of photon trajectories in the Kerr metric, starting very near the black hole event horizon from the geometrically thin accretion disk at the black hole equatorial plane and reaching a distant observer (a distant telescope). 
This reconstruction used the basic features of the tantalizing Blandford--Znajek mechanism for the heating of accretion plasma very near the black hole event horizon. Additionally, we neglected the absorption and dispersion of photons by the surrounding plasma. We call the resulting dark spot the ``event horizon silhouette''. In other words, we identified the dark regions in the EHT images with event horizon silhouettes. It must be especially stressed that the event horizon contour (silhouette) was projected at the celestial sphere inside the black hole shadow.

\section{Event Horizon Silhouette  $\equiv$   Lensed Image of the Black Hole Event Horizon}

\begin{figure}
	\centering
	\includegraphics[width=8.0cm]{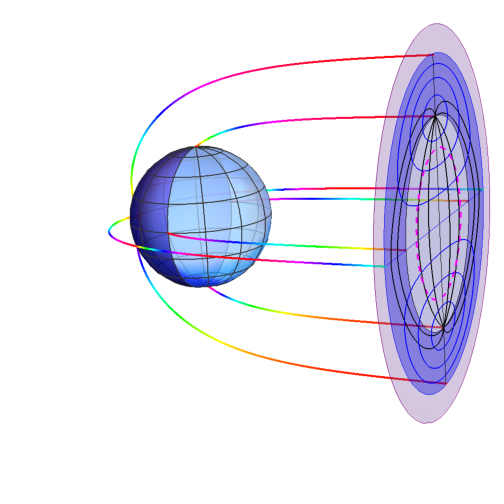} 
	\caption{Reconstruction of the Shwarzschild black hole event horizon silhouette using $3D$ trajectories of  photons (multicolored curves), which start very near the black hole event horizon and are registered by a distant observer (by a distant telescope). The event horizon silhouette always projects at the celestial sphere within the classical black hole shadow with a radius $3\sqrt{3}$. Meanwhile, the corresponding radius of event horizon silhouette is $r_{\rm h}\simeq4.457$ \cite{doknaz20,doknaz19}.}
	\label{fig1}
\end{figure}

In numerical calculations of test particle trajectories in the Kerr metric, it is useful to use the Boyer--Lindquist coordinate system  \cite{BoyerLindquist} with dimensional coordinates; for example, the dimensional radial length unit is \ $GM/c^2$. The corresponding event horizon radius of the Kerr black hole is $r_{\rm h}=1+\sqrt{1-a^2}$, with spin in the range $0\leq a\leq1$. All photon trajectories in our Figures were calculated by using both differential (\ref{A1})--(\ref{A4}) and integral (\ref{A9})--(\ref{A11}) forms of equations of motion in the Kerr metric. The form of a dark event horizon silhouette is numerically calculated from photon trajectories, which start near the event horizon and are registered by a distant observer (see details in Appendix~\ref{AppendixA}). 
In all our figures, the dashed red ring is an event horizon radius in the absence of gravity, while the red arrow is the direction of black hole rotation axes.  Figure~\ref{fig1} presents the reconstruction of a Shwarzschild black hole event horizon silhouette using $3D$ trajectories of  photons (multicolored curves), which start very near the black hole event horizon and are registered by a distant observer. (by a distant telescope).

\begin{figure}
	\centering
	\includegraphics[angle=0,width=0.42\textwidth]{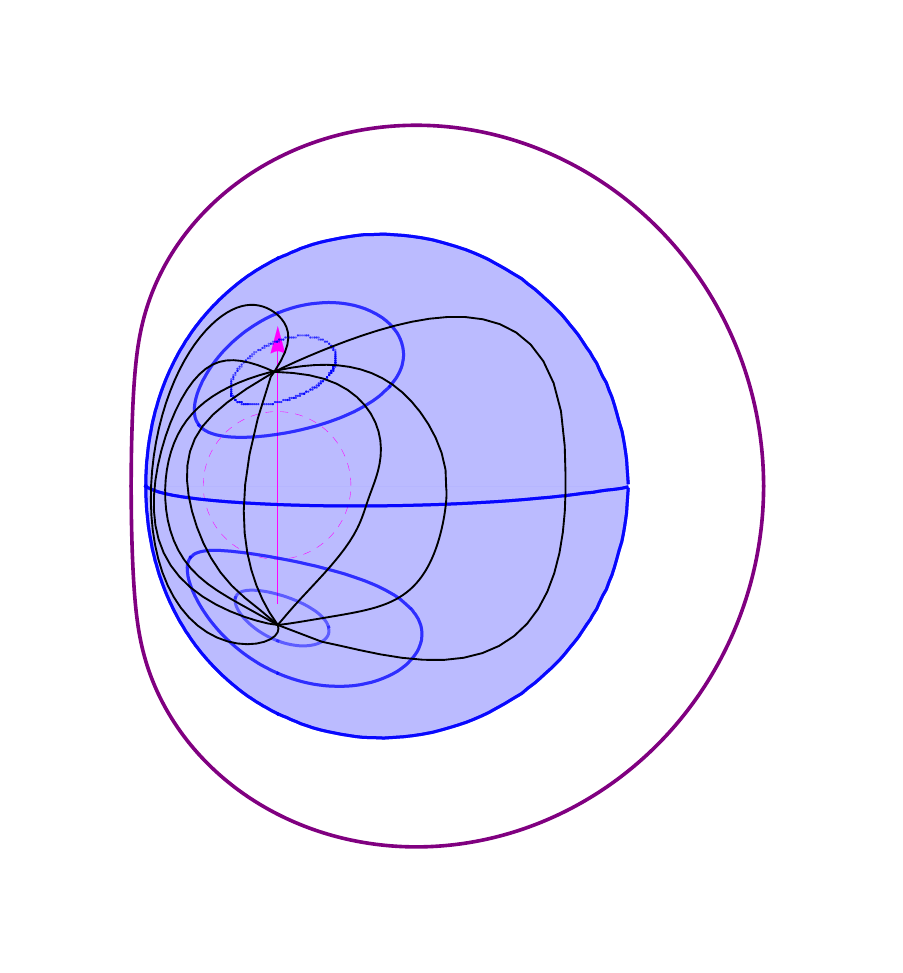} 	
	\hfill
	\includegraphics[angle=0,width=0.47\textwidth]{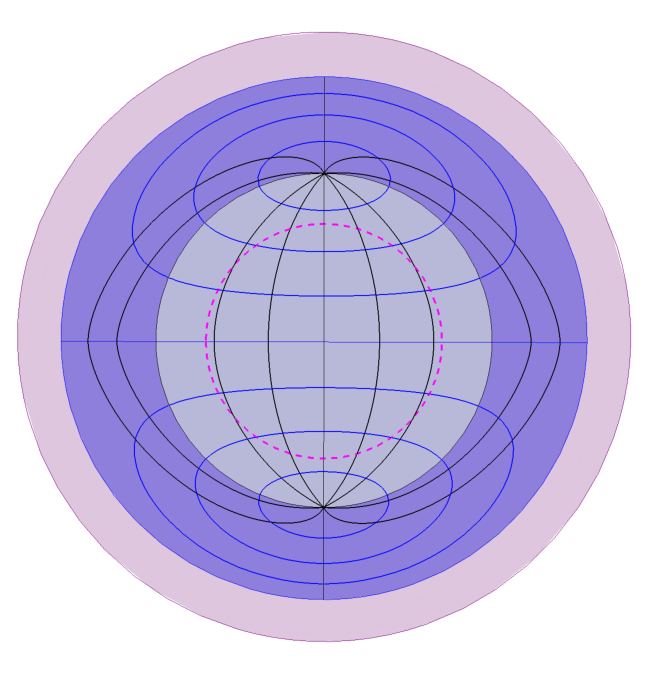} 
	\caption{The face profiles of the event horizon silhouettes in the Shwarzschild case (right panel, $a=0$) and extreme Kerr black hole (left panel, $a=1$). It is supposed that a distant observer is placed slightly above the black hole equatorial plane.}
	\label{fig2}
\end{figure}

Figure~\ref{fig2} demonstrates the face profiles of the event horizon silhouettes in the Shwarzschild case ($a=0$) and extreme Kerr black hole ($a=1$) corresponding to the SgrA* rotation axis inclination with respect to the line of sight of a distant observer. It is supposed that a distant observer is placed at the equatorial plane of the black hole. Note that the vertical sizes of shadows for this inclination of the black hole rotation axis are the same for all values of the spin parameter $0\leq a\leq1$.

\section{Forms and Sizes of Dark Spots in the Black Hole Images}

\begin{figure}[H]
	\centering
	\includegraphics[angle=0,width=10.5cm]{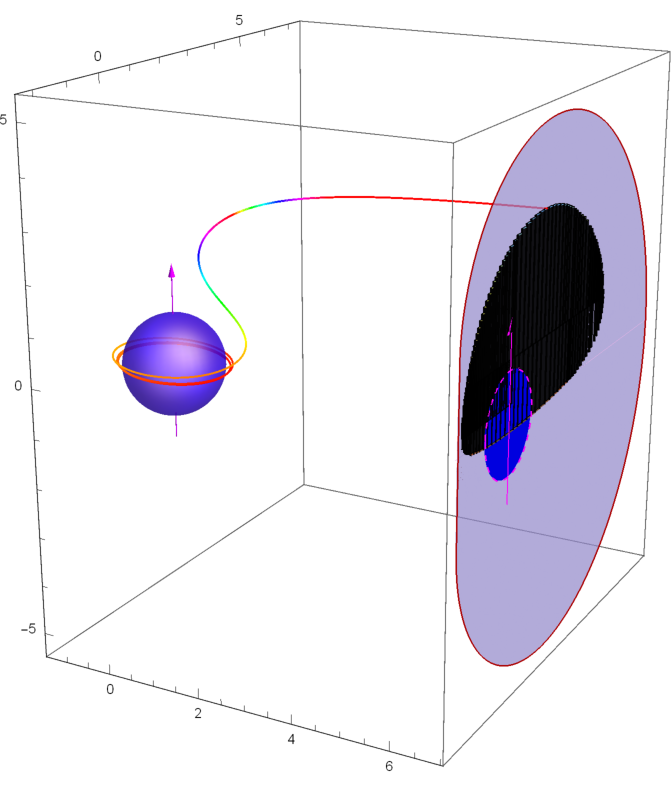} 
	\caption{A $3D$ picture of the SMBH SgrA* with the spin parameter $a=1$. It is supposed that a distant observer is placed slightly above the black hole equatorial plane at the polar angle $\theta_0=84^\circ$. The inner part of the dark spot is a lensed image of the north hemisphere of the event horizon globe. The outline of the dark spot is a lensed image of the event horizon equator. The multicolored curve is a numerically calculated trajectory of photon, starting from the radius $r=1.01r_h$ at the black hole equatorial plane and coming to a distant observer.}
	\label{fig3}
\end{figure}

The dark spot in the described black hole image is a locus of points with an absence of photons,  which start very near the black hole event horizon and are registered by a distant observer.

\begin{figure}[H]
	\centering
	\makebox{\includegraphics[width=0.23\textwidth,trim=0cm -2.5cm 0cm 0cm]{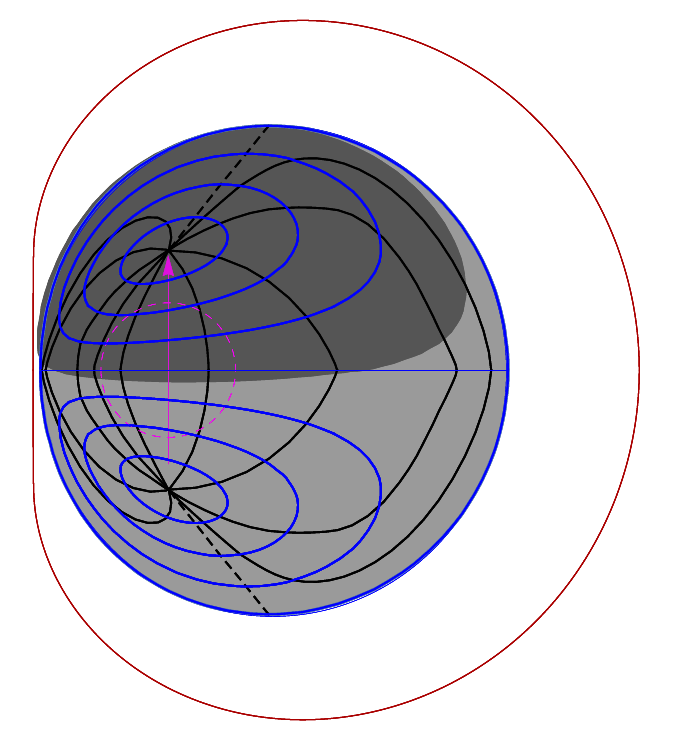}}
	\hfill
	\makebox{\includegraphics[width=0.37\textwidth]{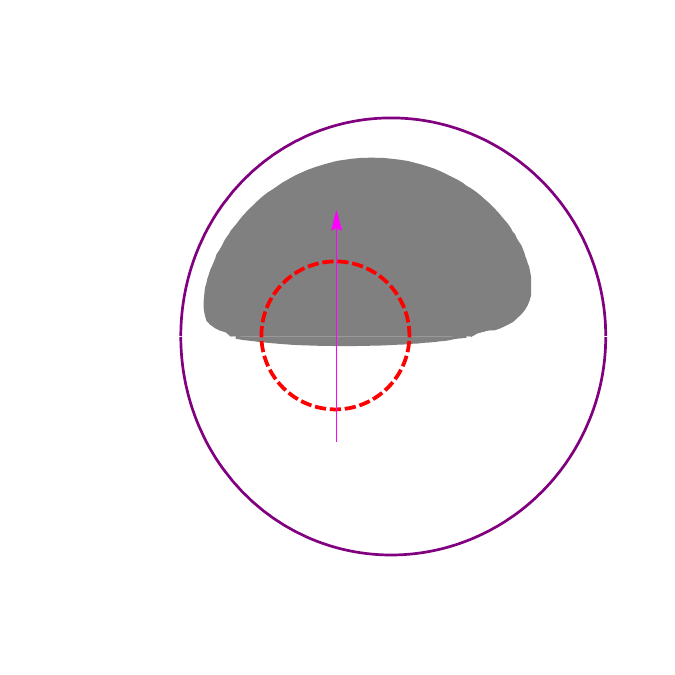}}	
	\makebox{\includegraphics[width=0.37\textwidth]{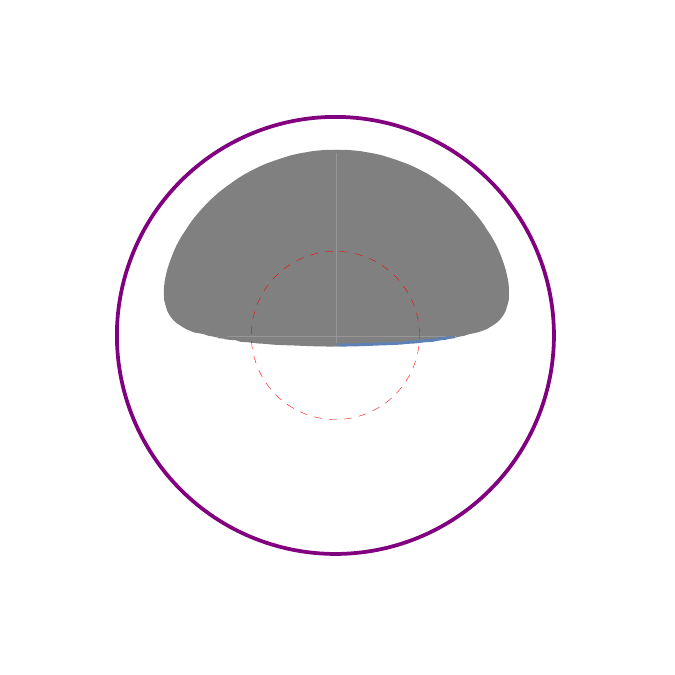}}
	\vspace{-16pt}
	\caption{Examples of dark spots for the case of SMBH SgrA* with different values of spin parameter $a$, obtained by numerical calculation of possible photon trajectories in the model of geometrically thin accretion disk: left panel, \ $a=1$, middle panel, \ $a=0.65$, right panel, \ $a=0$. It is supposed that a distant observer is placed slightly above the black hole equatorial plane at the polar angle $\theta_0=84^\circ$.}
	\label{fig4}
\end{figure}

\begin{figure}
	\centering
	\mbox{\includegraphics[width=0.48\textwidth]{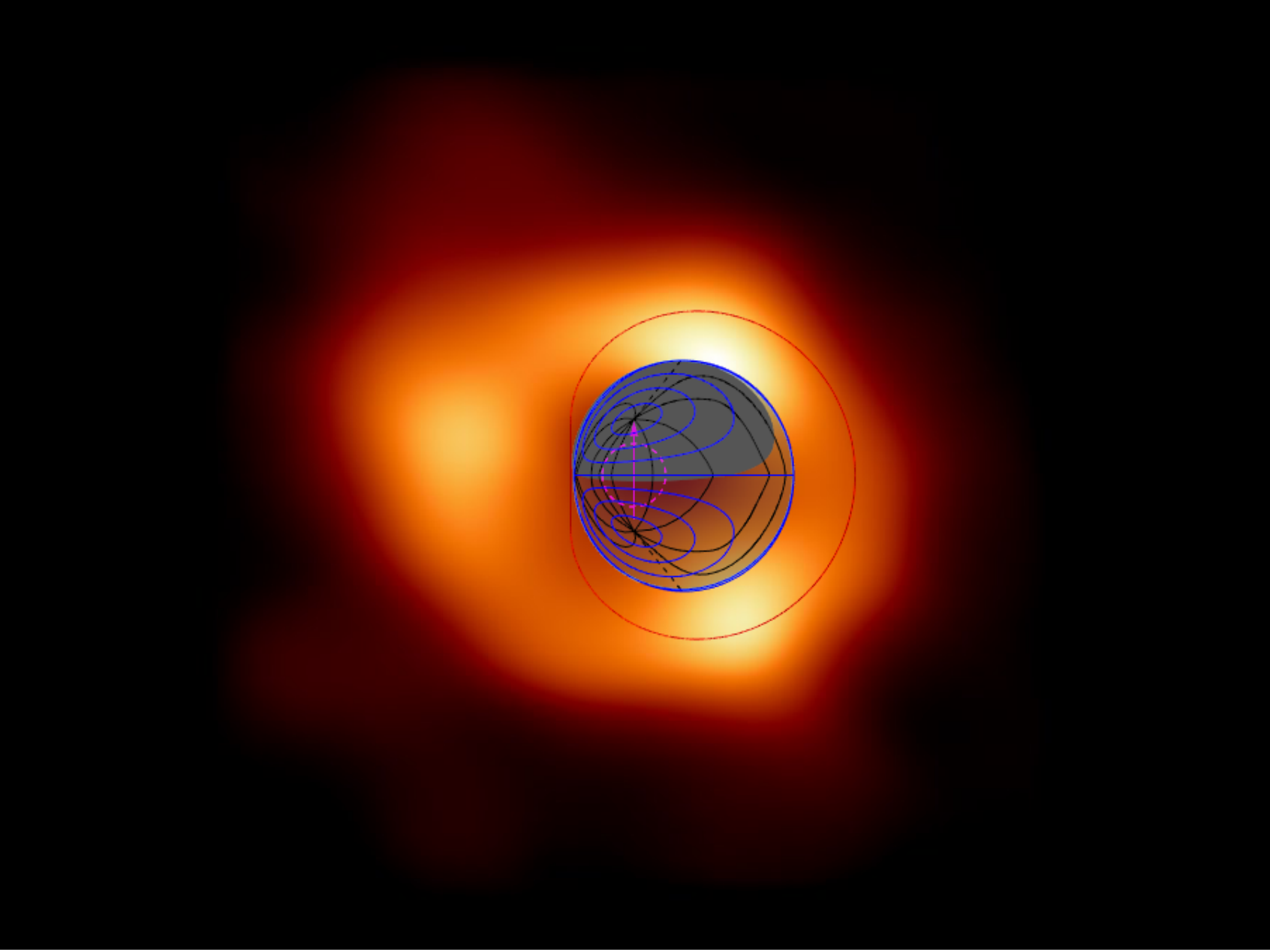}}
	\hfill
	\makebox{\includegraphics[width=0.48\textwidth]{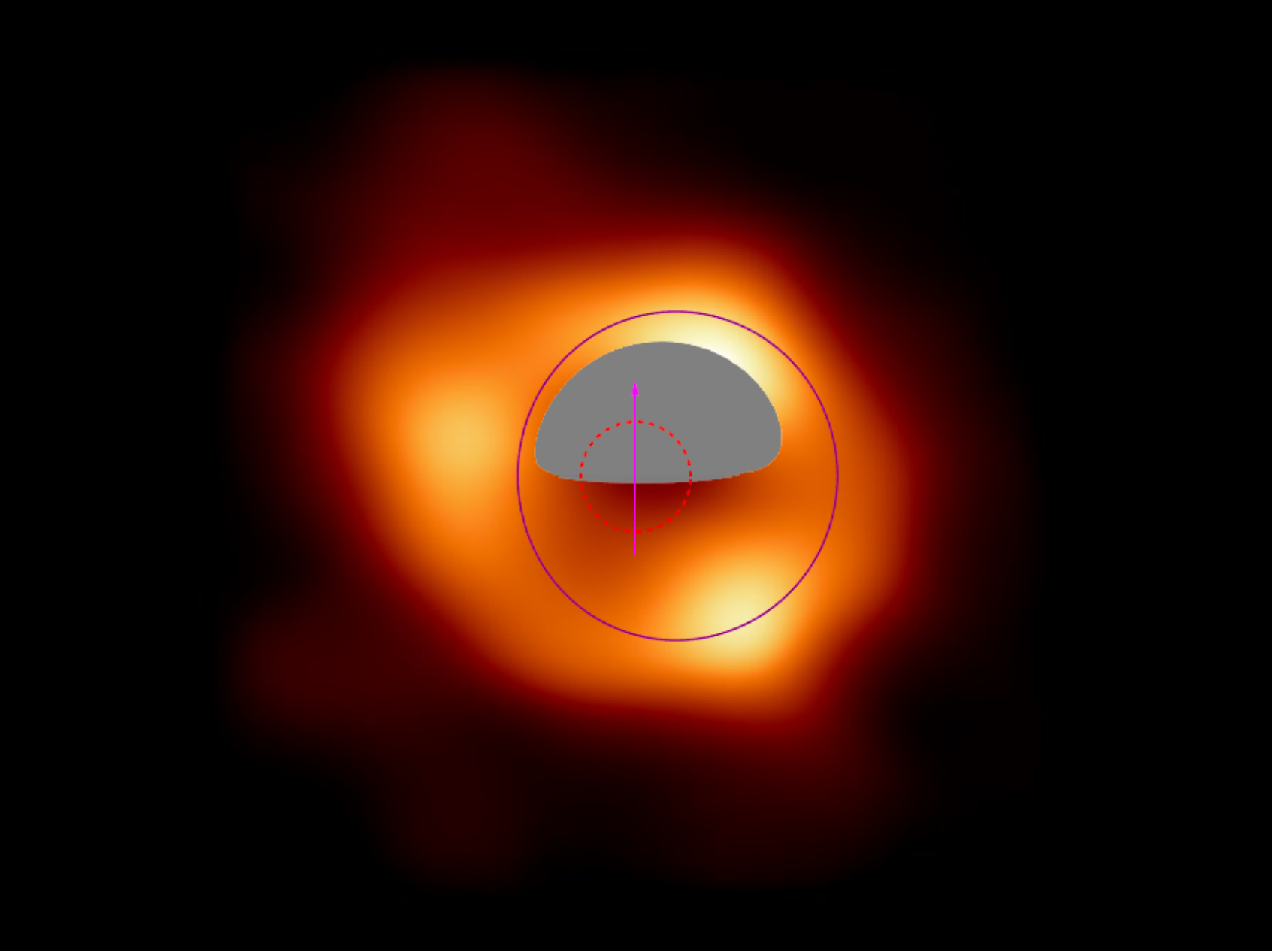}}
	\caption{Superposition of the modeled dark spot with the EHT image of SgrA*: left panel, \ $a=1$, \ right panel, \  $a=0.65$. The dark spots, which are the lensed images of event horizons, are always projected at the celestial sphere inside the awaited positions of the classical black hole shadows, which are invisible in both cases of SgrA* and M87*.}
	\label{fig5}
\end{figure}

Figure~\ref{fig3} shows a $3D$ picture of the SMBH SgrA* with the spin parameter $a=1$, where it is supposed that a distant observer is placed at the polar angle $\theta_0=84^\circ$. In this case, the dark image of the black hole corresponds to the north hemisphere of the event horizon globe. The boundary of this dark image is the event horizon equator. The multicolored curve is a numerically calculated trajectory of a photon with horizontal impact parameter $\lambda=-1.493$ and vertical impact parameter $q=3629$ (see Appendix~\ref{AppendixA} for definitions), starting from the radius $r=1.01r_{\rm h}$ at the black hole equatorial plane and coming to a distant observer. Figure~\ref{fig4} demonstrates some examples of dark spots for the case of SMBH SgrA* with different values of spin parameter $a$, obtained by a numerical calculation of possible photon trajectories in the model of a geometrically thin accretion disk. The superposition of the modeled dark spot with the EHT image of SgrA* is shown in Figure~\ref{fig5}.

\begin{figure}
	\centering
	\makebox{\includegraphics[width=0.32\textwidth,]{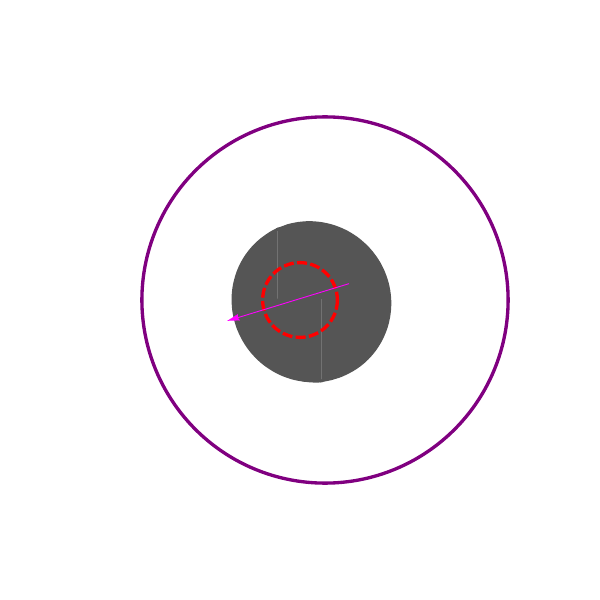}}
	\hfill
	\makebox{\includegraphics[width=0.32\textwidth]{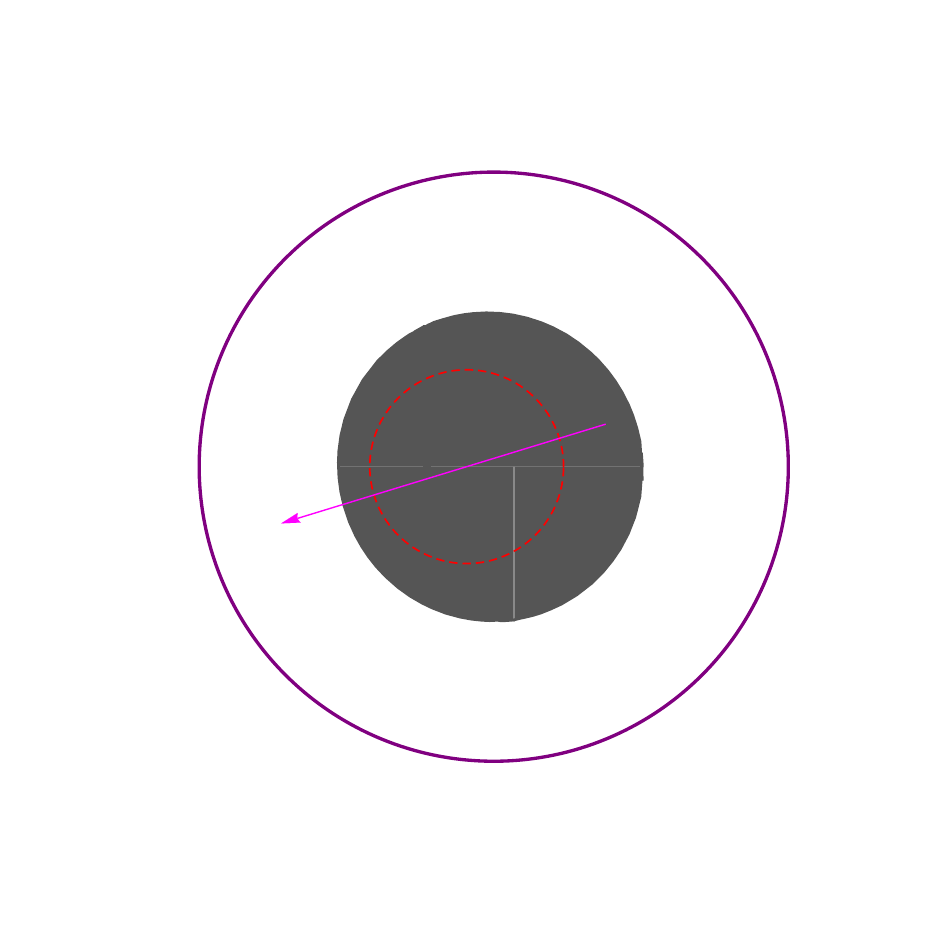}}	
		\hfill
	\makebox{\includegraphics[width=0.32\textwidth]{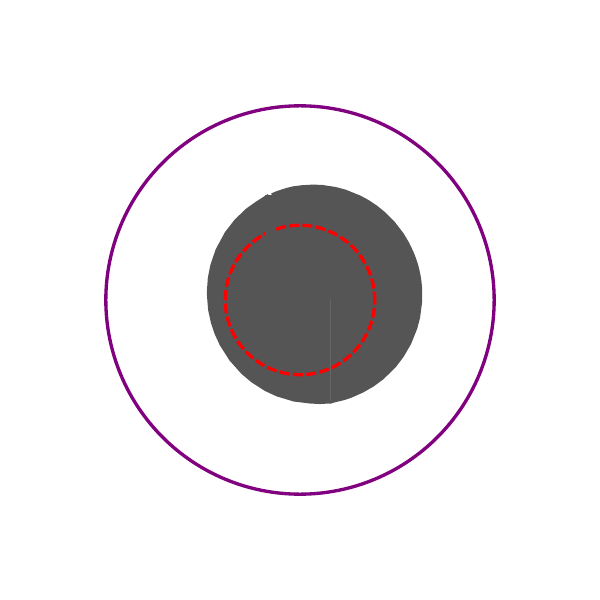}}
	\caption{Examples of dark spots for the case of SMBH M87* with a distant observer at the polar angle $\theta_0=84^\circ$: \ left panel, \ $a=1$; \ middle panel, \ $a=0.75$; \ right panel, \ $a=0$.}
	\label{fig6}
\end{figure}

A $17^\circ$  inclination angle of the M87* rotation axis with respect to a distant observer is supposed~\cite{Walker18,Nalewajko20}. Figure~\ref{fig6} demonstrates examples of dark spots for the case of SMBH M87* with a distant observer at the polar angle~$\theta_0=84^\circ$ and spin values~$a=1$, $a=0.75$, and \ $a=0$, respectively.

\begin{figure}[H]
	\centering
	\mbox{\includegraphics[width=0.48\textwidth]{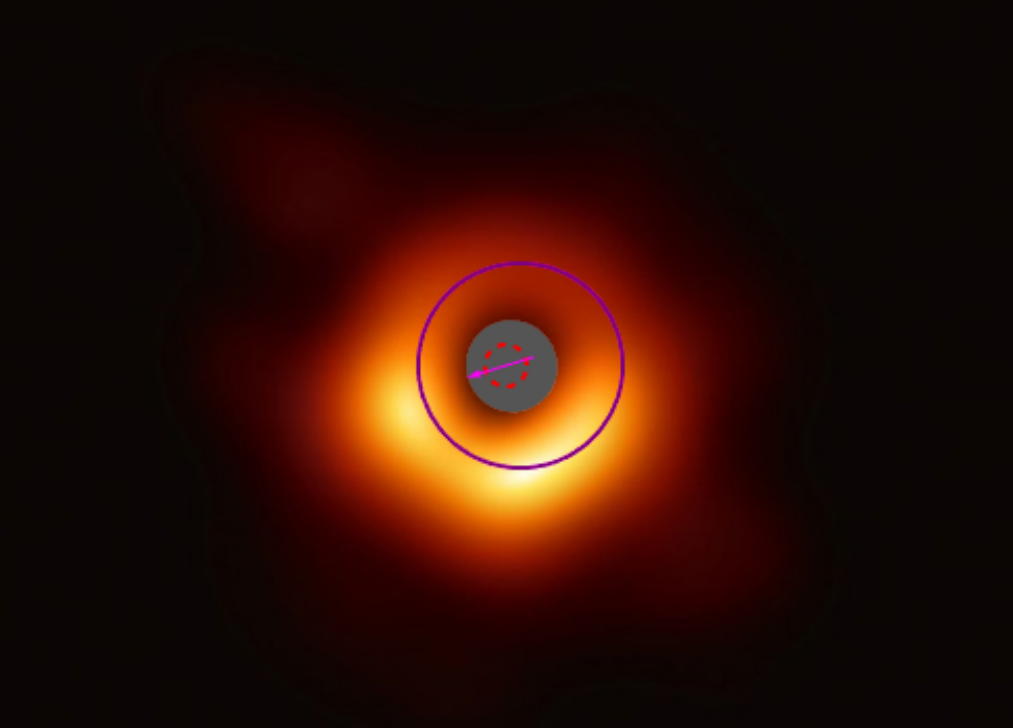}}
	\hfill
	\makebox{\includegraphics[width=0.48\textwidth]{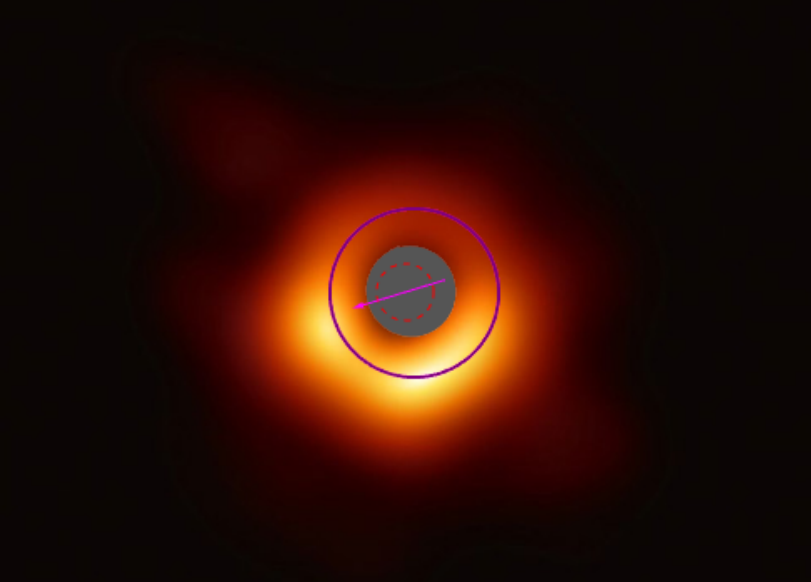}}
	\caption{Superposition of the modeled dark spot with the EHT image of M87*: Left panel, \ $a=1$; \ right panel, \ $a=0.75$. A $17^\circ$  inclination angle of the black hole rotation axis at the celestial sphere is supposed.}
	\label{fig7}
\end{figure}

\begin{figure}[H]
	\centering
	\includegraphics[angle=0,width=9cm]{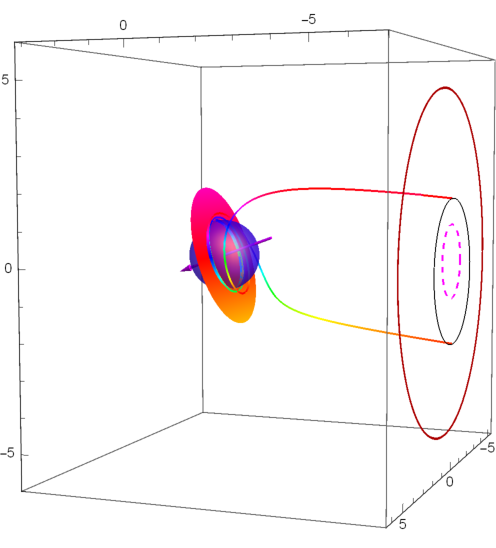} 
	\caption{A $3D$ picture of the SMBH M87* with spin parameter $a=1$ surrounded by thin accretion disk. A $17^\circ$ inclination angle of M87* rotation axis with respect to a distant observer is supposed.}
	\label{fig8}
\end{figure}

A superposition of the modeled dark spot with the EHT image of SgrA* is shown in  Figure~\ref{fig7}. Figure~\ref{fig8} demonstrates the $3D$ picture of the SMBH M87*. Here, a $17^\circ$ inclination angle of the M87* rotation axis with respect to a distant observer is supposed. 
A thin accretion disk is shown schematically by an ellipsoid with changing colors. The two multicolored curves are the trajectories of photons, which start from the accretion disk very near the black event horizon. The orbital parameters of these photons are, accordingly, $(\lambda=-0.0475,q= 2.194)$ and $(\lambda=-0.02869,q=1.518)$ \cite{doknaz21}. The purple curve is the classical black hole shadow, projected at the celestial sphere. In this case, the dark spot is the event horizon south hemisphere, while the outer boundary of the dark black hole image (the gray curve) is the gravitationally lensed event horizon equator.

\section{Spin Values of SgrA* and M87*}

In this paper, the dark spots in the EHT images of supermassive black holes SgrA* and M87* were modeled by supposing the existence of a thin accretion disk, which highlights the black hole in the very vicinity of its event horizon. 
We elucidate the applicability of this model by having in mind the Blandford--Znajek mechanism for heating the accretion plasma near the black hole event horizon. Additionally, the existence of hot accretion plasma very near the black hole event horizon is supported by numerical GRMHD simulations. 

By using numerical calculations of photon trajectories in the Kerr metric, we demonstrate that a dark spot in the black hole image is the image of its event horizon globe. There is one-to-one correspondence between points in the $3D$ event horizon globe and points in the $2D$ lensed event horizon image. In other words, the black hole is viewed from all sides at once. The lensed images of event horizons are always projected at the celestial sphere inside the awaited positions of the classical black hole shadows, which are invisible in both cases of M87* and SgrA*. 

We evaluated the spins of SMBHs SgrA* and M87* by using numerical calculations of the corresponding dark spots in the Kerr metric. In particular, we used trajectories of photons, which start very near the event horizon  and are registered by a distant observer. We used the sizes of dark spots in the images of SgrA* and M87* for inferring their spins, $0.65<a<0.9$ and $a>0.75$, accordingly, which are in general agreement with similar estimations by other methods. The confidence level of these spin estimations is $1\sigma$. These estimations may be improved or corrected during future observations with an advanced version of EHT.

\section{Conclusions}

Here, we evaluated the spins of SMBHs SgrA* and M87* by using the sizes of dark spots in their EHT images. We made numerical calculations of the corresponding dark spots by using trajectories of photons, which start very near the event horizon and are registered by a distant observer. 
We had in mind the Blandford--Znajek mechanism for heating the accretion plasma near the black hole event horizon. 

\begin{figure}[H]
		\centering
	\mbox{\includegraphics[width=0.48\textwidth]{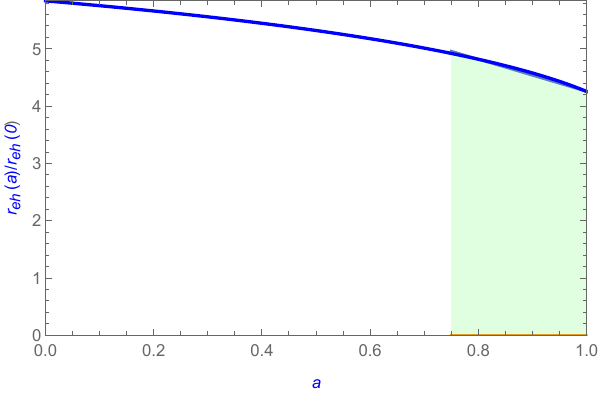}}
	\hfill
	\makebox{\includegraphics[width=0.48\textwidth]{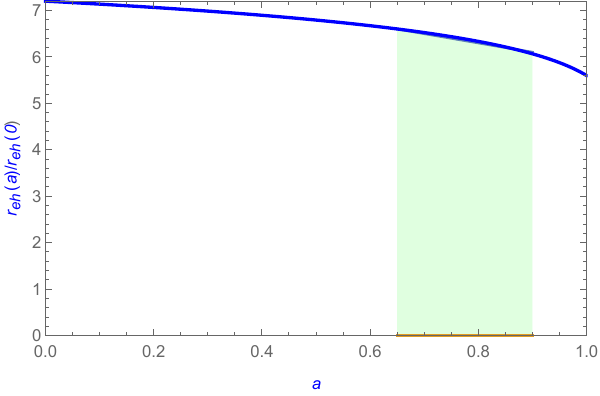}}
	\caption{Dependence of the dark spot size, $r_{\rm h}(a)$, on the black hole spin parameter $a$, respectively, for SgrA* (left panel) and M87* (right panel). The light green regions mark the possible range of the black hole spin parameters: $065\leq a\leq0.9$ for SgrA* and $0.75\leq a\leq1$ for M87*.}
	\label{fig9}
\end{figure}

The resulting dependence of the dark spot size, $r_{\rm h}(a)$, on the black hole spin parameter $a$ is shown in Figure~\ref{fig9}, respectively, for SgrA* and M87*. A comparison of the modeled sizes of dark spots in the images of black holes with similar ones in the EHT images reveals conclusively the possible range of spin parameters for SgrA*, $065\leq a\leq0.9$, and for M87*, $0.75\leq a\leq1$. Note that the values of spin parameters obtained here from the sizes of dark spots in the EHT images are in general agreement with similar estimations by other methods  \cite{Broderick08,Li09,Feng17,Sobyanin18,Nokhrina19,Tamburini19,Bambi19,Nemmen19,Davoudiasl19}.

We numerically calculated the possible forms and sizes of the dark spots in the images of SMBHs SgrA* and M87* by using the thin accretion disk model and supposing the Blandford--Znajek mechanism for heating the accretion plasma near the black hole event horizon. A comparison of modeled dark spots with similar ones in the EHT images reveals the possible values of spin parameters: $065\leq a\leq0.9$ for SgrA* and $0.75\leq a\leq1$ for M87*, respectively. The accuracy of these spin estimations is rather poor, with a confidence level of only $1\sigma$. The problem of the inevitable absorption and dispersion of photons, emitted by accretion matter and traversing toward a distant telescope, must also be noted. To overcome this problem, it is requested to construct telescopes for high-frequency photons. Future observations with the ``Advanced EHT'' or especially with the projected ``Millimetron Space Observatory'' may drastically increase the image sharpness of SMBHs SgrA* and~M87*.

\begin{acknowledgements}
	I am grateful to E.O. Babichev, V.A. Berezin, Yu.N. Eroshenko and N.O. Nazarova for stimulating discussions.
\end{acknowledgements}

\appendix
\section{Geodesics in Kerr Metric}
\label{AppendixA}

We used the classical Brandon Carter equations of motion for test particles in the Kerr metric with the famous integrals of motion (see  \cite{Carter68,BoyerLindquist,BPT,CunnBardeen72,CunnBardeen73,mtw,Chandra,Galtsov} for details):
\begin{eqnarray} 
	\Sigma\frac{dr}{d\tau} &=& \pm \sqrt{R(r)}, 
	\label{A1} \\
	\Sigma\frac{d\theta}{d\tau} &=& \pm\sqrt{\Theta(\theta)}, \label{A2} \\
	\Sigma\frac{d\phi}{d\tau} &=& L\sin^{-2}\theta+a(\Delta^{-1}P-E), 
	\label{A3}	\\
	\Sigma\frac{dt}{d\tau} &=& a(L-aE\sin^{2}\theta)+(r^2+a^2)\Delta^{-1}P.
	\label{A4}	
\end{eqnarray}

In these equations, $\tau$ is a proper time (for $\mu\neq0$) or an affine parameter of particle (for $\mu=0$), the effective radial potential $R(r)$ in (\ref{A1}) is
\begin{equation}	\label{A5} 
	R(r) = P^2-\Delta[\mu^2r^2+(L-aE)^2+Q], \quad 	P=E(r^2+a^2)-a L,
\end{equation}
and the effective polar (latitudinal) potential $\Theta(\theta)$ in (\ref{A1}), which controls the motion in the polar (latitude) direction, is
\begin{equation}	\label{A6} 
	\Theta(\theta) = Q-\cos^2\theta[a^2(\mu^2-E^2)+L^2\sin^{-2}\theta].
\end{equation}

As usual, dimensionless orbital parameters were used, defining the motion of massive test particles, $\gamma=E/\mu$, $\lambda=L/E$, and $q^2=Q/E^2$, respectively. The motion of massless particles (like photons) depends on only two parameters, $\lambda=L/E$ and $q^2=Q/E^2$. These parameters are related with the horizontal and vertical impact parameters, $\alpha$ and $\beta$, viewed by the distant observer at the celestial sphere (see, e.g.,  \cite{Bardeen73,CunnBardeen73} for more details),
\begin{equation}
	\alpha=-\frac{\lambda}{\sin\theta_0}, \quad \beta=\pm\sqrt{\Theta(\theta_0)}.
	\label{A7} 
\end{equation}

\section{Classical black hole shadow}
\label{AppendixB}

The outer boundary of the classical black hole shadow in the Kerr metric may be determined by the joint resolution of equations for photon spheres, $R(r)=0$  and $[rR(r)]'=0$ (see, e.g.,  \cite{Bardeen73,Chandra} for more details). The corresponding mutual solution of these two equations may be written in parametric form $(\lambda,q)=(\lambda(r),q(r))$ if the black hole is viewed from its equatorial plane,
\begin{equation} \label{A10}
	\lambda=\frac{-r^3+3r^2-a^2(r+1)}{a(r-1)}, \quad
	q^2=\frac{r^3[4a^2-r(r-3)^2]}{a^2(r-1)^2}.
\end{equation}

\begin{figure}[H]
	\centering
	\mbox{\includegraphics[width=0.49\textwidth]{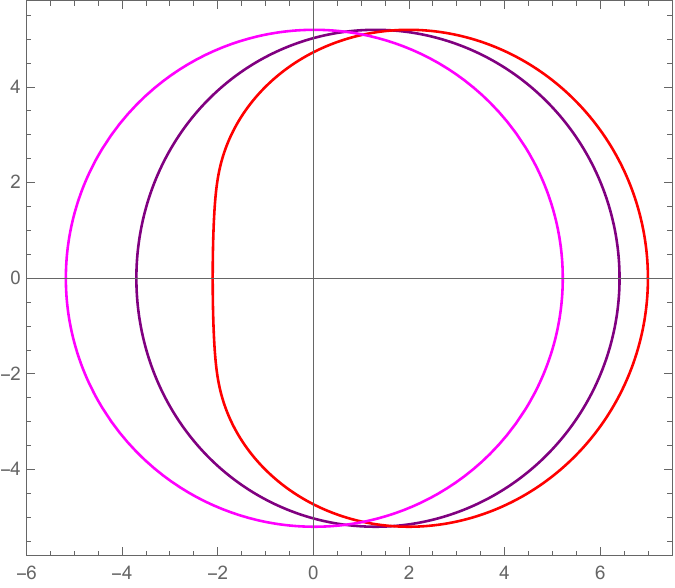}}
	\hfill
	\makebox{\includegraphics[width=0.44\textwidth]{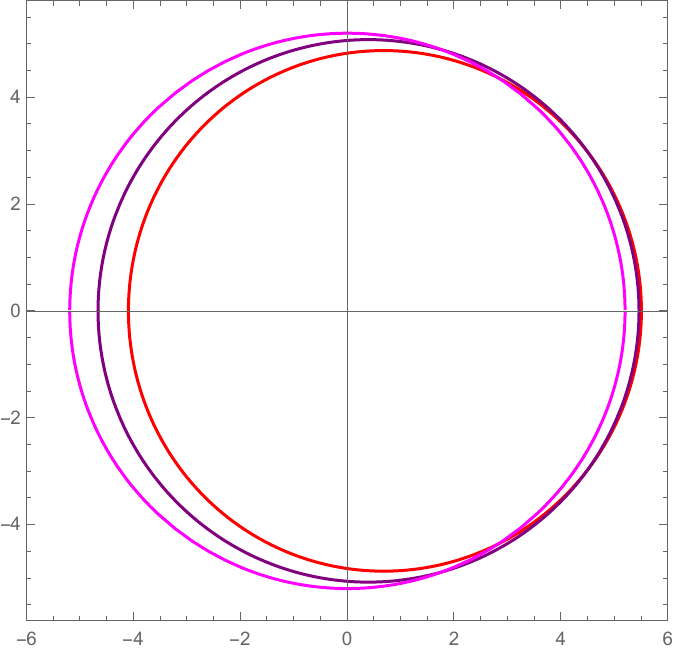}}	\caption{Classical black hole shadows, projected at the celestial sphere, for different spin parameters $a$ in the Kerr metric and possible inclinations of SMBH rotation axes with respect to the polar angle $\theta_0$ of a distant observer: SgrA* in the left panel ($\theta_0=90^\circ$, \ $a=1$ \ (red), \ $a=0.65$ \ (purple), \ $a=0$ \ (magenta)) and M87* in the right panel\ ($\theta_0=163^\circ$, \ $a=1$ \ (red), \ $a=0.75$ \ (purple), $a=0$ (magenta)).}
	\label{fig10}
\end{figure}

Examples of the classical black hole shadows, projected at the celestial sphere, are shown in Figure~\ref{fig10} for different spin parameters $a$ in the Kerr metric and possible inclinations of SMBHs M87* and SgrA* rotation axes at the celestial sphere.

\section{Brandon Carter equations of motion in the Kerr metric in the integral form}
\label{AppendixC}

In our calculations, we used the standard integral form of Brandon Carter equations of motion in the Kerr metric \cite{Carter68,Chandra}:
\begin{equation}\label{A9}
	\fint^r\frac{dr}{\sqrt{R(r)}}
	=\fint^\theta\frac{d\theta}{\sqrt{\Theta(\theta)}},
\end{equation}
\begin{equation}
	\varphi=\!\fint^r\!\!\frac{a(r^2+a^2-\lambda a)}{(r^2\!-\!2r\!+\!a^2)\sqrt{R(r)}}\,dr
	\!+\!\fint^\theta\!\frac{\lambda\!-\!a\sin^2\theta}{\sin^2\theta\sqrt{\Theta(\theta)}}\,d\theta,
	\label{A10} 
\end{equation}
\begin{equation}
	t=\fint^r\!\!\frac{(r^2+a^2)P}{(r^2\!-\!2r\!+\!a^2)\sqrt{R(r)}}\,dr
	+\fint^\theta\frac{(L\!
		-\!aE\sin^2\theta)a}{\sqrt{\Theta(\theta)}}\,d\theta.
	\label{A11} 
\end{equation}

The integrals in (\ref{A2})--(\ref{A11}) are the path integrals along the curved trajectory. A major specific feature of these integrals is the changing of the sign at each turning point, both in the radial and polar directions. As a result, these path integrals monotonically grow along the curved particle trajectory. In the absence of turning points, these path integrals become ordinary ones. For example, the path integral in (\ref{A9}) can be written in the absence of turning points through the ordinary integral as
\begin{equation}\label{A15}
	\int^{r_0}_{r_s}\frac{dr}{\sqrt{R(r)}}
	=\int^{\theta_0}_{\theta_s}\frac{d\theta}{\sqrt{\Theta(\theta)}}.
\end{equation}

In this equation, $r_s$ and $\theta_s$ are the radial and polar starting points of the particle, respectively, and  $r_0$ and $\theta_0$ are the finishing points along the curved trajectory. 

In particular, the photon trajectories in Figures~\ref{fig1}--\ref{fig6} were calculated by using solutions of Equations  (\ref{A9})--(\ref{A11}) with a starting radial coordinate $r_s=1.01r_{\rm h}$, with a finishing radial coordinate $r_0=10^5r_{\rm h}$, with different starting polar coordinates $\theta_s$ in the range $0\leq\theta_s\leq\pi$, and, respectively, with different azimuth coordinates $\varphi_s$ in the range $0\leq\varphi_s\leq2\pi$ (see details in  \cite{Dokuch22,doknaz21,doknaz20,Dokuch19,doknaz19,doknazsm19}).

The second example is with one turning point in the polar direction, $\theta_{\rm min}(\lambda,q)$, (an extremum of latitudinal potential $\Theta(\theta)$). The corresponding  integral Equation (\ref{A9}) can be written in this case through the ordinary integrals as
\begin{equation}\label{A116}
	\int^{r_0}_{r_s}\frac{dr}{\sqrt{R(r)}}
	=\int^{\theta_s}_{\theta_{\rm min}}\frac{d\theta}{\sqrt{\Theta(\theta)}}
	+\int^{\theta_0}_{\theta_{\rm min}}\frac{d\theta}{\sqrt{\Theta(\theta)}}.
\end{equation}

The third specific case is the particle trajectory with one turning point in the polar direction, $\theta_{\rm min}(\lambda,q)$, and one turning point in the radial direction, $r_{\rm min}(\lambda,q)$. The path integral in  (\ref{A9}) in this case can be written  through the ordinary integrals as 
\begin{equation}\label{eq24c}
	\int_{r_{\rm min}}^{r_s}\!\!\frac{dr}{\sqrt{R(r)}}
	+\int_{r_{\rm min}}^{r_0}\!\!\frac{dr}{\sqrt{R(r)}}
	=\!\!\int_{\theta_{\rm min}}^{\theta_s}\!\!\frac{d\theta}{\sqrt{\Theta(\theta)}}
	+\!\!\int_{\theta_{\rm min}}^{\theta_0}\!\!\frac{d\theta}{\sqrt{\Theta(\theta)}}.
\end{equation}

Figures~\ref{fig1}--\ref{fig7} show the results of numerical calculations for these path integrals.


\begin{thebibliography}{999}
		
\bibitem{EHT1} 		Akiyama, K.; Alberdi, A.; Alef, W.; Asada, K.; Azulay, R.; Baczko, A.-K.; Ball, D.; Balokovic, M.; Barrett, J.; Bintley, D.;~et~al. M87 Event Horizon Telescope Results. I. The Shadow of the Supermassive Black Hole. {\em Astrophys. J.} {\bf 2019}, {\em 875}, L1.
		
		
		\bibitem{EHT2}
		Akiyama, K.; Alberdi, A.; Alef, W.; Asada, K.; Azulay,~R.; Baczko, A.K.; Ball, D.; Balokovic, M.; Barrett, J.; Bintley, D.;~et~al. M87 Event Horizon Telescope Results. II. Array and Instrumentation. {\em Astrophys. J.} {\bf 2019}, {\em 875}, L2.
		
		
		\bibitem{EHT3} Akiyama, K.; Alberdi, A.; Alef, W.; Asada, K.; Azulay, R.; Baczko, A.K.; Ball, D.; Balokovic, M.; Barrett, J.; Bintley, D.;~et~al. M87 Event Horizon Telescope Results. III. Data Processing and Calibration. {\em Astrophys. J.} {\bf 2019}, {\em 875}, L3.
		
		
		\bibitem{EHT4} Akiyama, K.; Alberdi, A.; Alef, W.; Asada, K.; Azulay, R.; Baczko, A.K.; Ball, D.; Balokovic, M.; Barrett, J.; Bintley, D.;~et~al. M87 Event Horizon Telescope Results. IV. Imaging the Central Supermassive Black Hole. {\em Astrophys. J.} {\bf 2019}, {\em 875}, L4.
		
		
		\bibitem{EHT5} Akiyama, K.; Alberdi, A.; Alef, W.; Asada, K.; Azulay, R.; Baczko, A.K.; Ball, D.; Balokovic, M.; Barrett,~J.; Bintley, D.;~et~al. M87 Event Horizon Telescope Results. V. Physical Origin of the Asymmetric Ring. {\em Astrophys. J.} {\bf 2019}, {\em 875}, L5.
		
		
		\bibitem{EHT6} Akiyama, K.; Alberdi, A.; Alef, W.; Asada, K.; Azulay, R.; Baczko,~A.-K.; Ball, D.; Balokovic, M.; Barrett, J.; Bintley, D.; ~et~al. M87 Event Horizon Telescope Results. VI. The Shadow and Mass of the Central Black Hole. {\em Astrophys. J.} {\bf 2019}, {\em 875}, L6.
		
		
		\bibitem{EHT1S} {Akiyama, K.; Alberdi, A.; Alef, W.; Algaba, J.C.; Anantua, R.; Asada, K.; Azulay, R.; Bach, U.; Baczko, A.K.; Ball, D.;~et~al.} First Sagittarius A* Event Horizon Telescope Results. I. The Shadow of the Supermassive Black Hole in the Center of the Milky Way. \emph{Astrophys. J.} \textbf{2022}, \emph{930}, L12.
		
		
		\bibitem{EHT2S}{Akiyama, K.; Alberdi, A.; Alef, W.; Algaba, J.C.; Anantua, R.; Asada, K.; Azulay, R.; Bach, U.; Baczko, A.K.; Ball, D.;~et~al.} First Sagittarius A* Event Horizon Telescope Results. II. EHT and Multiwavelength Observations, Data Processing, and Calibration. \emph{Astrophys. J.} \textbf{2022}, \emph{930}, L13.
		
		
		\bibitem{EHT3S}{Akiyama, K.; Alberdi, A.; Alef, W.; 
			Algaba, J.C.; Anantua, R.; Asada, K.; Azulay, R.; Bach, U.; Baczko, A.K.; Ball, D.;~et~al.} First Sagittarius A* Event Horizon Telescope Results. III. Imaging of the Galactic Center Supermassive Black Hole. \emph{Astrophys. J.} \textbf{2022}, \emph{930}, L14.  
		
		
		\bibitem{EHT4S}{Akiyama, K.; Alberdi, A.; Alef, W.; Algaba, J.C.; 	Anantua, R.; Asada, K.; Azulay, R.; Bach, U.; Baczko, A.K.; Ball, D.;~et~al.} First Sagittarius A* Event Horizon Telescope Results. IV. Variability, Morphology, and Black Hole Mass. \emph{Astrophys. J.} \textbf{2022}, \emph{930}, L15.
		
		
		\bibitem{EHT5S}{Akiyama, K.; Alberdi, A.; Alef, W.; Algaba, J.C.; Anantua, R.; Asada, K.; Azulay, R.; Bach, U.; Baczko, A.K.; Ball, D.;~et~al.} First Sagittarius A* Event Horizon Telescope Results. V. Testing Astrophysical Models of the Galactic Center Black Hole. \emph{Astrophys. J.} \textbf{2022}, \emph{930}, L16.
		
		
		\bibitem{EHT6S}{Akiyama, K.; Alberdi, A.; Alef, W.; Algaba, J.C.; Anantua, R.; Asada, K.; Azulay, R.; Bach, U.; Baczko, A.K.; Ball, D.;~et~al.} First Sagittarius A* Event Horizon Telescope Results. VI. Testing the Black Hole Metric. \emph{Astrophys. J.} \textbf{2022}, \emph{930}, L17.
		
		
		\bibitem{Bardeen73} Bardeen, J.M. Timeline and Null Geodesics in the Kerr Metric. In {\em Black Holes}; DeWitt, C., DeWitt, B.S., Eds.; Gordon and Breach: New York, NY, USA, 1973; pp.~217--239. 
		
		
		\bibitem{Falcke00} Falcke, H.; Melia, F.; Agol, E. Viewing the Shadow of the Black Hole at the Galactic Center. {\em Astrophys. J.} {\bf 2000}, {\em 528}, L13--L16.
		
		
		\bibitem{ZakhPaoIngrNuc05} Zakharov, A.F.; De Paolis, F.; Ingrosso G.; Nucita, A.A. Measuring the black hole parameters in the galactic center with RADIOASTRON. {\em New Astron.} {\bf 2005}, {\em 10}, 479--489.
		
		
		\bibitem{JohPsaltis10} Johannsen, T.; Psaltis, D. Testing the No-hair Theorem with Observations in the Electromagnetic Spectrum. II. Black Hole Images. {\em Astrophys. J.} {\bf 2010}, {\em 718}, 446--454.
		
		
		\bibitem{Grenzebach14}  Grenzebach, A.; Perlick, V.; L\"ammerzahl, C. Photon regions and shadows of Kerr-Newman-NUT black holes with a cosmological constant. {\it Phys. Rev. D} {\bf 2014}, {\em 89}, 124004.
		
		
		\bibitem{Grenzebach15} Grenzebach, A.; Perlick, V.; L\"ammerzahl, C. Photon regions and shadows of accelerated black holes. {\it Int. J. Mod. Phys. D}  {\bf 2015}, {\em 24}, 1542024.
		
		
		\bibitem{Cunha18a} Cunha, P.V.P.; Herdeiro, C.A.R. Shadows and strong gravitational lensing: A brief review.  {\it Gen. Relativ. Gravit.} {\bf 2018}, {\em 50}, 42.
		
		
		\bibitem{Cunha18b} Cunha, P.V.P.; Herdeiro, C.A.R.; Rodriguez, M.J. Does the black hole shadow probe the event horizon geometry? {\it Phys. Rev. D} {\bf 2018}, {\em 97},084020.
		
		
		\bibitem{Huang18} Huang, Y.; Dong, Y.-P.; Liu, D.-J. Revisiting the shadow of a black hole in the presence of a plasma. {\it Int. J. Mod. Phys. D} {\bf 2018}, {\em 27},1850114.
		
		\bibitem{BisnovatyiTsupko18} Bisnovatyi-Kogan, G.S.; Tsupko O.Yu. Shadow of a black hole at cosmological distances. {\it Phys. Rev. D} {\bf 2018}, {\em 98}, 084020.
		
		\bibitem{Gralla15} Gralla, S.E.; Porfyriadis, A.P.; Warburton, N. Particle on the innermost stable circular orbit of a rapidly spinning black hole.  \emph{Phys. Rev. D} {\bf 2015}, {\em 92}, 064029.
		
		
		\bibitem{Strom16} Gralla, S.E.; Lupsasca, A.; Strominger, A. Near-horizon Kerr magnetosphere. {\it Phys. Rev. D} {\bf 2016}, {\em 93}, 104041.
		
		
		\bibitem{Gralla16} Gralla, S.E.; Zimmerman, A.; Zimmerman, P. Transient instability of rapidly rotating black holes. {\it Phys. Rev. D} {\bf 2016}, {\em 94}, 084017.
		
		
		\bibitem{Strom17} Porfyriadis, A.P.; Shi, Y.; Strominger, A. Photon emission near extreme Kerr black holes. {\it Phys. Rev. D} {\bf 2017}, {\em 95}, 064009.
		
		
		\bibitem{Strom18} Gralla, S.E.;  Lupsasca, A.; Strominger, A.  Observational signature of high spin at the Event Horizon Telescope. {\it Mon. Not. R. Astron. Soc.}  {\bf 2017}, {\em 475}, 3829--3853.
		
		\bibitem{Brenneman} Brenneman, L. {\em Measuring the Angular Momentum of Supermassive Black Holes}; Springer: New York, NY, USA, 2013.
		
		
		\bibitem{Reynolds} Reynolds, C.S. Observing black holes spin. {\it Nat. Astron.} {\bf 2019}, {\em 3}, 41--47. 
		
		\bibitem{Tchekhovskoy11} Tchekhovskoy, A.; Narayan, R.; McKinney, J.C. Efficient generation of jets from magnetically arrested accretion on a rapidly spinning black hole. {\it Mon. Not. R. Astron. Soc.}  {\bf 2011}, {\em 418}, L79--L83.
		
		
		\bibitem{Tchekhovskoy12} Tchekhovskoy, A.; McKinney, J.C.; Narayan, R. General Relativistic Modeling of Magnetized Jets from accretion Black Holes. {\it J. Phys. Conf. Ser.}  {\bf 2012}, {\em 372}, 012040.
		
		
		\bibitem{Tchekhovskoy12b} McKinney, J.C.; Tchekhovskoy A.; Blandford R.D. General relativistic magnetohydrodynamic simulations of magnetically choked accretion flows around black holes. {\em Mon. Not. R. Astr. Soc.} {\bf 2012}, \emph{423}, 3083--3117.
		
		
		\bibitem{Tchekhovskoy15} Ressler, S.M.; Tchekhovskoy, A.; Quataert, E.; Chandra, M.;  Gammie, C.F. Electron thermodynamics in GRMHD simulations of low-luminosity black hole accretion.  {\em Mon. Not. R. Astr. Soc.} {\bf 2015}, \emph{454}, 1848--1870. 
		
		
		\bibitem{Tchekhovskoy17} Ressler, S.M.; Tchekhovskoy, A.; Quataert, E.; Gammie, C.F. The disc-jet symbiosis emerges: Modelling the emission of Sagittarius A* with electron thermodynamics. {\em Mon. Not. R. Astr. Soc.} {\bf 2017}, \emph{467}, 3604--3619.
		
		
		\bibitem{Tchekhovskoy17b} Foucart, F.; Chandra, M.; Gammie, C.F.; Quataert, E.; Tchekhovskoy, A. How important is non-ideal physics in simulations of sub-Eddington accretion on to spinning black holes? {\em Mon. Not. R. Astr. Soc.} {\bf 2017}, \emph{470}, 2240--2252.
		
		
		\bibitem{Ryan18} Ryan, B.R.; Ressler, S.M.; Dolence, J.C.; Gammie, C.F.; Quataert, E. Two-Temperature GRRMHD Simulations of M87.  {\it Astrophys. J.} {\bf 2018}, {\em 864}, 126, 13 pp.
		
		
		\bibitem{BlandfordZ} Blandford, R.D.; Znajek, R.L. Electromagnetic extraction of energy from Kerr black holes. {\em Mon. Not. R. Astr. Soc.} {\bf 1977}, \emph{179}, 433.
		
		
		\bibitem{Luminet79} Luminet, J.-P. Image of a spherical black hole with thin accretion disk. {\em Astron. Astrophys.} {\bf 1979}, {\em 75}, 228--235.
		
		
		\bibitem{Dexter09} Dexter, J.; Agol, E.; Fragile, P.C. Millimeter Flares and VLBI Visibilities from Relativistic Simulations of Magnetized Accretion onto the Galactic Center Black Hole. {\it Astrophys. J.} {\bf 2009}, {\em 703}, L142--L146.
		
		
		\bibitem{Bromley97} Bromley, B.C.; Chen, K.; Miller, W.A. Line Emission from an Accretion Disk around a Rotating Black Hole: Toward a Measurement of Frame Dragging. {\it Astrophys. J.} {\bf 1997}, {\em 475}, 57--64.
		
		
		\bibitem{Fanton97} Fanton, C.; Calvani, M.; de Felice, F.; Cadez, A. Detecting Accretion Disks in Active Galactic Nuclei. {\it Publ. Astron. Soc. Jpn.} {\bf 1997}, {\em 49}, 159--169.
		
		
		\bibitem{Armitage03} Armitage, P.J.; Reynolds C.S. Silhouette of a Dressed Black Hole.  {\it MNRAS} {\bf 2003}, {\em 341},  1041--1050.
		
		
		\bibitem{Fukue03} Fukue, J. Silhouette of a Dressed Black Hole.  {\it Publ. Astron. Soc. Jpn.} {\bf 2003}, {\em 55}, 155--159.
		
		
		\bibitem{Fukue03b} Fukue, J. Light-Curve Diagnosis of a Hot Spot for Accretion-Disk Models.  {\it Publ. Astron. Soc. Jpn.} {\bf 2003}, {\em 55}, 1121--1125.
		
		
		\bibitem{Tamburini11} Tamburini, F.; Thid\'e, B.; Molina-Terriza, G.; Anzolin, G. Twisting of light around rotating black holes. {\it Nat. Phys.} {\bf 2011},  {\em 7}, 195--197.
		
		
		\bibitem{Ru-SenLu16} Lu, R.; Roelofs, F.; Fish, V.L.; Shiokawa, H.; Doeleman, S.S.; Gammie, C.F.; Falcke, H.; Krichbaum, T.P.; Zensus, J.A. Imaging an Event Horizon: Mitigation of Source Variability of Sagittarius A*. {\it Astrophys. J.} {\bf 2016}, {\em 817}, 173.
		
		
		\bibitem{Luminet19} Luminet, J.-P. An Illustrated History of Black Hole Imaging: Personal Recollections (1972--2002). {\it arXiv} {\bf 2019}, arXiv:1902.11196.
		
		
		\bibitem{Dokuch22}  Dokuchaev, V.I. Physical origin of the dark spot at the image of supermassive black hole SgrA* revealed by the EHT collaboration. {\it Astronomy} {\bf 2022}, {\em 1}, 93--98.
		
		
		\bibitem{doknaz21} Dokuchaev, V.I.; Nazarova, N.O. Modeling the motion of a bright spot in jets from black holes M87* and SgrA*. {\it Gen. Relativ. Gravit.} {\bf 2021}, {\em 53}, 83.
		
		
		\bibitem{doknaz20} Dokuchaev, V.I.; Nazarova, N.O. Silhouettes of invisible black holes.  {\it Physics-Uspekhi} {\bf 2020}, {\em 63}, 583--600.
		
		
		\bibitem{doknaz19} Dokuchaev, V.I.; Nazarova, N.O. Event horizon image within black hole shadow. {\it J. Exp. Theor. Phys.} {\bf 2019}, {\em 128}, 578--585.
		
		
		\bibitem{Dokuch19} Dokuchaev, V.I. To see invisible: Image of the event horizon within the black hole shadow. {\it Int. J. Mod. Phys. D} {\bf 2019}, {\em 28}, 1941005.
		
		
		\bibitem{doknazsm19} Dokuchaev, V.I.; Nazarova, N.O.; Smirnov, V.P. Event horizon silhouette: Implications to supermassive black holes M87* and SgrA*. {\it Gen. Relativ. Gravit.} {\bf 2019}  {\em 51}, 81. 
		
		\bibitem{BoyerLindquist} Boyer, R.H.; Lindquist, R.W. Maximal Analytic Extension of the Kerr Metric. {\it J. Math. Phys.} {\bf 1967}, {\em 8}, 265--282.
		
		
		\bibitem{Walker18} Walker, R.C.; Hardee, P.E.; 	Davies, F.B.; Ly, C.; Junor, W. The structure and dynamics of the sub-parsec scale jet in M87 based on 50 VLBA observations over 17 years at 43 GHz. {\it Astrophys. J.} {\bf 2018}, {\em 855}, 128.
		

		
		\bibitem{Nalewajko20} Nalewajko, K.; Sikora, M.,; R\'oz\`ansk\`a, A. On the orientation of the crescent image of M87*. {\em Astron. Astrophys.} 
		{\bf 2020}, \emph{634}, 7. 
		
		
		
		\bibitem{Broderick08} Broderick, A.E.; Loeb, A. Imaging the Black Hole Silhouette of M87: Implications for Jet Formation and Black Hole Spin.  {\it Astrophys J.} {\bf 2009}, {\em 697}, 1164--1179.
		
		
		\bibitem{Li09} Li, Y.; Yuan, Y.; Wang, J.; Wang, J.; Zhang, S. Constraining spins of supermassive black holes from TeV variability. II. fully general relativistic calculations.  {\it Astrophys. J.} {\bf 2009}, {\em 699}, 513--524.
		
		
		\bibitem{Feng17} Feng, J.; Wu, Q. Constraint on the black hole spin of M87 from the accretion-jet model. {\it Mon. Not. R. Astron. Soc.}  {\bf 2017}, {\em 470}, 612--616.
		
		
		\bibitem{Sobyanin18} Sob'yanin, D.N. Black hole spin from wobbling and rotation of the M87 jet and a sign of a magnetically arrested disc. {\it Mon. Not. R. Astron. Soc.}  {\bf 2018}, {\em 479}, L65--L69.
		
		
		\bibitem{Nokhrina19} Nokhrina, E.E.; Gurvits, L.I.; Beskin, V.S.; Nakamura, M.; Asada, K.; Hada, K. M87 black hole mass and spin estimate through the position of the jet boundary shape break. {\it arXiv} {\bf 2019}, arXiv:1904.05665.
		
		
		\bibitem{Tamburini19} Tamburini, F; Thid\'e, B.; Della Valle, M. Measurement of the spin of the M87 black hole from its observed twisted light. {\it arXiv} {\bf 2019}, arXiv:1904.07923.
		
		
		\bibitem{Bambi19} Bambi, C.; Freese, K.; Vagnozzi, S.; Visinelli, L. Testing the rotational nature of the supermassive object M87* from the circularity and size of its first image.  {\it arXiv} {\bf 2019}, arXiv:1904.12983.
		
		
		\bibitem{Nemmen19} Nemmen, R. The Spin of M87*. {\it arXiv} {\bf 2019}, arXiv:1905.02143.
		
		
		\bibitem{Davoudiasl19} Davoudiasl, H.; Denton, P.B. Ultralight Boson Dark Matter and Event Horizon Telescope Observations of M87. {\it Phys. Rev. Lett.}  {\bf 2019},  {\em 123}, 021102.
		
		
		
		\bibitem{Carter68} Carter, B. Global Structure of the Kerr Family of Gravitational Fields. {\it Phys. Rev.} {\bf 1968}, {\em 174}, 1559--1571.
		
		
		\bibitem{BPT} Bardeen, J.M.; Press, W.H.; Teukolsky, S.A. Rotating Black Holes: Locally Nonrotating Frames, Energy Extraction, and Scalar Synchrotron Radiation. {\it Astrophys. J.} {\bf 1972}, {\em 178}, 347--370.
		
		
		\bibitem{CunnBardeen72} Cunningham, C.T.; Bardeen, J.M. 		The Optical Appearance of a Star Orbiting an Extreme Kerr Black Hole. {\em Astrophys. J.} {\bf 1972}, {\em 173}, L137–L142.
		
		
		\bibitem{CunnBardeen73} Cunningham, C.T.; Bardeen, J.M. 
		The Optical Appearance of a Star Orbiting an Extreme Kerr Black Hole. {\it Astrophys. J.} {\bf 1973}, {\em 183}, 237--264.
		
		
		\bibitem{mtw} Misner, C.W.; Thorne, K.S.; Wheeler, J.A. {\em Gravitation}; W. H. Freeman: San Francisco, CA, USA, 1973.
		
		
\bibitem{Chandra} Chandrasekhar, S. {\it The Mathematical Theory of Black Holes}; Clarendon Press: Oxford, UK, 1983.

		
		\bibitem{Galtsov} Gal'tsov, D.V. {\em Particles and Fields in the Vicinity of Black Holes}; Moscow Univ. Press: Moscow, Russia, 1986. (In~Russian)

\end{thebibliography}
\end{document}